\documentclass[twocol]{ametsoc}
\usepackage{dsfont}

\journal{jas}

\bibpunct{(}{)}{;}{a}{}{,}

\title{The linear response function of an idealized atmosphere. Part 2: \\ Implications for the practical use of the Fluctuation-Dissipation Theorem and the role of operator's non-normality}

    \authors{Pedram Hassanzadeh\correspondingauthor{Pedram Hassanzadeh, 24 Oxford Street, Cambridge, MA 02138.}}
    \affiliation{Department of Earth and Planetary Sciences and Center for the Environment, Harvard University, Cambridge, Massachusetts}
    \email{hassanzadeh@fas.harvard.edu}

    \extraauthor{Zhiming Kuang}
    \extraaffil{\small Department of Earth and Planetary Sciences and John A. Paulson School of Engineering and Applied Sciences, Harvard University, Cambridge, Massachusetts}

\abstract{A linear response function (LRF) relates the mean-response of a nonlinear system to weak external forcings and vice versa. Even for simple models of the general circulation, such as the dry dynamical core,  the LRF cannot be calculated from first principles due to the lack of a complete theory for eddy-mean flow feedbacks. According to the Fluctuation-Dissipation Theorem (FDT), the LRF can be calculated using only the covariance and lag-covariance matrices of the unforced system. However, efforts in calculating the LRFs for GCMs using FDT have produced mixed results, and the reason(s) behind the poor performance of the FDT remains unclear. In Part~1 of this study, the LRF of an idealized GCM, the dry dynamical core with Held-Suarez physics, is accurately calculated using Green's functions. In this paper (Part~2), the LRF of the same model is computed using FDT, which is found to perform poorly for some of the test cases. The accurate LRF of Part~1 is used with a linear stochastic equation to show that dimension-reduction by projecting the data onto leading EOFs, which is commonly used for FDT, can alone be a significant source of error. Simplified equations and examples of $2 \times 2$ matrices are then used to demonstrate that this error arises because of the non-normality of the operator. These results suggest that errors caused by dimension-reduction are a major, if not the main, contributor to the poor performance of the LRF calculated using FDT, and that further investigations of dimension-reduction strategies with a focus on non-normality are needed.} 

\begin{document}

\maketitle

%
\section{Introduction} \label{sec:fdt}
In statistical physics, the Fluctuation-Dissipation Theorem (FDT) relates the mean-response of a nonlinear system to weak external forcing with the internal variability of the system \citep[see][for a review]{marconi2008fluctuation}. According to FDT, the system's linear response function (LRF), ${\pmb{\mathsf{L}}}$, which relates the mean-response $\langle {\mathrm{\mathbf{x}}}  \rangle$ to an imposed (external) forcing ${\mathrm{\mathbf{f}}}$ via 
\begin{eqnarray}
{\pmb{\mathsf{L}}} \langle {\mathrm{\mathbf{x}}}  \rangle = - \langle {\mathrm{\mathbf{f}}}  \rangle 
\label{eq:steady}
\end{eqnarray}
can be calculated using only the covariance and lag-covariances of the unforced system, i.e., when ${\mathrm{\mathbf{f}}}=0$ and the system fluctuates because of its internal dynamics (details are discussed in section~\ref{sec:fdtformula}). Here $\langle \cdot \rangle$ means long-time averaged and ${\mathrm{\mathbf{x}}}$ is the state-vector response, i.e., deviation from the time-mean state-vector of the unforced system.     

\citet{leith1975climate} introduced the FDT to the climate science, argued that the climate system approximately satisfies the conditions for the theorem to hold, and formulated how the LRF in Eq.~(\ref{eq:steady}) can be calculated using the fluctuations of the unforced system. Since then many studies have used climate models of varying degrees of complexity to examine the LRFs calculated using different forms of the FDT and the implications of this theorem \citep[e.g.,][]{bell1980climate,north1993fluctuation,cionni2004fluctuation,gritsun2007climate,abramov2007blended,gritsun2008climate,ring08,gerber2008annular,gerber2008testing,kirk2009diagnosis,majda2010low,gershgorin2010test,achatz2013fluctuation,lutsko2015applying,fuchs2015exploration}. However, the competency of the FDT for the climate system remains unclear as some of these studies found FDT to only work qualitatively, although some other, such as \citet{gritsun2007climate} and  \citet{fuchs2015exploration} found promising quantitative skills. 

Whether the reported failure of the FDT is because of the invalidity of the underlying assumptions for the climate system, or because of practical problems (e.g., associated with using limited datasets), or a combination of both, is unclear (see section~\ref{sec:fdtformula} for further discussions). How accurately the FDT holds for the climate system is important, not only because of the possibility to construct skillful LRFs from unforced GCM simulations or even ambitiously, from observational records, but also because some of the implications of this theorem. For example, FDT relates the amplitude of the forced response to the timescales of internal modes of variability of the system \citep[e.g.,][]{ring08,shepherd2014}. Given that idealized and comprehensive GCMs overestimate the persistence of the leading mode of the extratropical variability in both hemispheres, i.e., the Annular Modes, by factors as large as $2-3$ (and even larger in some cases), an important implication of the FDT is that these models also overestimate the mean-response to external forcings by such factors, which has significant consequences for the climate sensitivity (see \citet{shepherd2014,gerber2008annular,gerber2008testing,ring08}; but also see \citet{simpson2016revisiting}).  

In Part~1 \citep{pedram16}, for zonally-averaged flows in the context of an idealized dry atmosphere, we derive and discuss Eq.~(\ref{eq:steady}) and show that the state-vector $\overline{\mathrm{\mathbf{x}}}$ can be reduced to $\overline{\mathrm{\mathbf{y}}} \equiv (\overline{u},\overline{T})$, where $\overline{u}$ and $\overline{T}$ are zonal-mean zonal-wind and temperature. We also calculate the LRF of an idealized GCM, the dry dynamical core with Held-Suarez physics, using Green's functions: we apply hundreds of weak localized forcings $\langle \overline{\mathrm{\mathbf{f}}}  \rangle $ (of $\overline{u}$ and $\overline{T}$), one at a time, to the GCM and calculate the mean-responses $\overline{\mathrm{\mathbf{y}}}$, which are then used along with the applied forcings to find the LRF through matrix inversion. To be consistent with Part~1, the LRF calculated for this idealized setup will be denoted with $\tilde{\pmb{\mathsf{M}}}$:
\begin{eqnarray}
\tilde{\pmb{\mathsf{M}}} \langle \overline{\mathrm{\mathbf{y}}}  \rangle = - \langle \overline{\mathrm{\mathbf{f}}}  \rangle 
\label{eq:steady2}
\end{eqnarray}  
where $\langle \overline{\mathrm{\mathbf{f}}}  \rangle$ is zonal-momentum and/or thermal forcing. Several tests in Part~1 show that the LRF calculated using Green's functions, $\tilde{\pmb{\mathsf{M}}}_\mathrm{GRF}$ hereafter, accurately reproducesthe mean-response to imposed thermal/mechanical forcings and vice versa. 

The goal of the current paper (Part~2) is to use the same idealized setup and $\tilde{\pmb{\mathsf{M}}}_\mathrm{GRF}$ to investigate why the LRF calculated using the FDT, $\tilde{\pmb{\mathsf{M}}}_\mathrm{FDT}$ hereafter, performs poorly in some cases. The paper is structured as follows. In section~\ref{sec:fdtformula} the formulation of the FDT is presented, in section~\ref{sec:model} the idealized GCM and calculation of $\tilde{\pmb{\mathsf{M}}}_\mathrm{FDT}$ from the simulations are briefly described, and in section~\ref{sec:fdttests} the performance of $\tilde{\pmb{\mathsf{M}}}_\mathrm{FDT}$ for a few test cases (same as Tests~1-3 in Part~1) is discussed. In sections~\ref{sec:fdtSDE} and \ref{sec:fdt2x2} we use linear stochastic equations with, respectively, $\tilde{\pmb{\mathsf{M}}}_\mathrm{GRF}$ and $2 \times 2$ matrices to show how the non-normality of the operator can affect the performance of the FDT. The paper is summarized in section~\ref{sec:sum}.                  

\section{Formulation of the FDT} \label{sec:fdtformula}
According to the most common formulation, the so-called quasi-Gaussian FDT, the LRF $\tilde{\pmb{\mathsf{M}}}_\mathrm{FDT}$ that relates the  mean-response $\langle \overline{\mathrm{\mathbf{y}}}  \rangle$ to an imposed forcing $\langle \overline{\mathrm{\mathbf{f}}}  \rangle$ via Eq.~(\ref{eq:steady2}) can be calculated as  
\begin{eqnarray}
\tilde{\pmb{\mathsf{M}}}_\mathrm{FDT}  = - \left[ \int_0^\infty  \pmb{\mathsf{C}}(\tau) \pmb{\mathsf{C}}(0)^{-1} \, d\tau \right]^{-1}
\label{eq:FDT}
\end{eqnarray}
where $\pmb{\mathsf{C}}(\tau)=\langle \overline{\mathrm{\mathbf{y}}}(\tau) \overline{\mathrm{\mathbf{y}}}(0)^\dag \rangle$ is the lag-$\tau$ covariance matrix ($\dag$ denotes the adjoint). Recently, \citet{majda2005information} and \citet{gritsun2007climate} have demonstrated that Eq.~(\ref{eq:FDT}) can be derived under conditions that are more closely satisfied by the atmosphere compared to those used by \citet{leith1975climate} and \citet{kraichnan1959classical}. Still, an important assumption involved in (\ref{eq:FDT}) is that $\overline{\mathrm{\mathbf{y}}}$ has Gaussian statistics; however, non-Gaussianity in dynamical and thermodynamic variables has been  found in observational data \citep[e.g.,][]{ruff2012long,huybers2014us,loikith15} and GCM simulations \citep[e.g.,][]{berner2007linear,franzke2007origin,sardeshmukh2009reconciling,pedram14}            

Furthermore, there are practical problems with calculating $\tilde{\pmb{\mathsf{M}}}_\mathrm{FDT}$ in Eq.~(\ref{eq:FDT}) generally due to the limited length
of the dataset. Realistically, the upper bound of the integral is replaced with a finite number $\tau_\infty$ and while a small $\tau_\infty$ degrades the approximation of the integral, a large $\tau_\infty$ can lead to an imprecise $\tilde{\pmb{\mathsf{M}}}_\mathrm{FDT}$ because of inaccuracies in $\pmb{\mathsf{C}}$ at large $\tau$ due to limited sample size. Additionally, the calculation of $\tilde{\pmb{\mathsf{M}}}_\mathrm{FDT}$ or $\tilde{\pmb{\mathsf{M}}}^{-1}_\mathrm{FDT}$ involves the inverse of the sum of the lag-covariance matrices or ${\pmb{\mathsf{C}}}(0)^{-1}$. These matrices can be close to singular because of short datasets and anisotropic internal fluctuations, which together result in the phase-space not being entirely excited by the fluctuations. The common remedy for this problem is to calculate $\pmb{\mathsf{C}}$ and $\tilde{\pmb{\mathsf{M}}}_\mathrm{FDT}$ on reduced dimensions, e.g., by first projecting the data onto a specified number ($n_\mathrm{EOF}$) of the leading Empirical Orthogonal Functions (EOFs) \citep{penland1989random,gritsun2007climate,ring08}. Another practical issue is the number of variables that are included in the state-vector $\overline{\mathrm{\mathbf{y}}}$. Some studies have used one variable such as zonal-wind or temperature and some other have used two or more variables. 

It is plausible that the reported inaccuracies in FDT and discrepancy in the previous studies are due to the invalidity of the assumptions underlying Eq.~(\ref{eq:FDT}), such as non-Gaussianity in the data, and/or practical problems such as short datasets and uncertainties in choosing $\tau_\infty$, $n_\mathrm{EOF}$, and $\overline{\mathrm{\mathbf{y}}}$. Several interesting studies have recently attempted to systematically address the issues related to sample size and dimension-reduction in calculations of $\pmb{\mathsf{C}}$ \citep{fuchs2015exploration,lutsko2015applying,cooper2013estimation}, state-vector reduction \citep{majda2010low}, and non-Gaussianity \citep{cooper2011climate,nicolis2015fluctuation}; however, further work is evidently needed to fully utilize the FDT for practical applications.

To further evaluate the performance of LRFs calculated from FDT and to better understand the potential sources of their inaccuracies, in section~\ref{sec:model} we have employed multivariate FDT and a long dataset to compute $\tilde{\pmb{\mathsf{M}}}_\mathrm{FDT}$ for the idealized GCM and in section~\ref{sec:fdttests} we have tested the performance of $\tilde{\pmb{\mathsf{M}}}_\mathrm{FDT}$ using Tests~1-3 of Part~1. 


\section{The Idealized Simulations and Calculation of $\tilde{\pmb{\mathsf{M}}}_\mathrm{FDT}$} \label{sec:model}
The model and setup are identical to Part~1. Briefly, we use the GFDL dry dynamical core, which is a pseudo-spectral GCM that solves the primitive equations on sigma levels. The GCM is used with the Held-Suarez setup \citep{held1994proposal}: the model is forced by Newtonian relaxation of temperature to a prescribed equinoctial radiative-equilibrium state with a specified equator-to-pole surface temperature difference, and Rayleigh drag with a prescribed rate is used to remove momentum from the low levels and $\nabla^8$ hyper-diffusion is used to remove enstrophy at small scales. The forcings, dissipations, and boundary conditions are all zonally-symmetric and symmetric between the two hemispheres. A T63 spectral resolution with $40$ equally-spaced sigma levels and $15$~min time-steps are used to solve the equations. Using the control-run setup of the model, which is identical to the control-run of Part~1 where all parameters are the same as in \citet{held1994proposal}, we have constructed $\tilde{\pmb{\mathsf{M}}}_\mathrm{FDT}$ from Eq.~(\ref{eq:FDT}) using a one million-day dataset as described below. 

An ensemble of $10$ simulations with the control-run setup, each $50000$~days, are used to create the employed dataset, which contains daily-averaged anomalous $\overline{u}$ and $\overline{T}$ of the last $49500$~days of each simulations where each variable is weighted by $\sqrt{\cos{\mu}}$ (where $\mu$ is the latitude) and normalized by area-averaged standard deviation of each pressure level (we did not find the performance of FDT sensitive to the weighting details). Covariance matrices are then calculated for each hemisphere of each simulation from $(\overline{u},\overline{T})$ (stacked together). The $20$ covariance matrices are subsequently averaged to calculate the EOFs of the ensemble. Then for a given $(\tau_\infty,n_\mathrm{EOF})$, the weighted daily-averaged anomalous $(\overline{u},\overline{T})$ is projected onto the first $n_\mathrm{EOF}$ EOFs using least-square linear regression, and the results are used to calculate the reduced-dimension covariance and lag-covariance matrices for each hemisphere in each simulation, which are then averaged for each $\tau$ to find $\pmb{\mathsf{C}}(\tau)$ for $\tau=0,1,2, \dots \tau_\infty$~days. The integral in (\ref{eq:FDT}) is evaluated using the trapezoidal rule.

To find the best performance of $\tilde{\pmb{\mathsf{M}}}_\mathrm{FDT}$, for each test we have tried $\tau_\infty=20,30,45,60,$ and $90$~days and $n_\mathrm{EOF}=64\,(97.5\%),113\,(99.0\%),165\,(99.5\%),200\,(99.7\%),$ and $300\,(99.8\%)$ where the parentheses show the explained variance



\section{Tests~1-3 for $\tilde{\pmb{\mathsf{M}}}_\mathrm{FDT}$} \label{sec:fdttests}
\begin{figure*}[t]
\centerline{\includegraphics[width=1\textwidth]{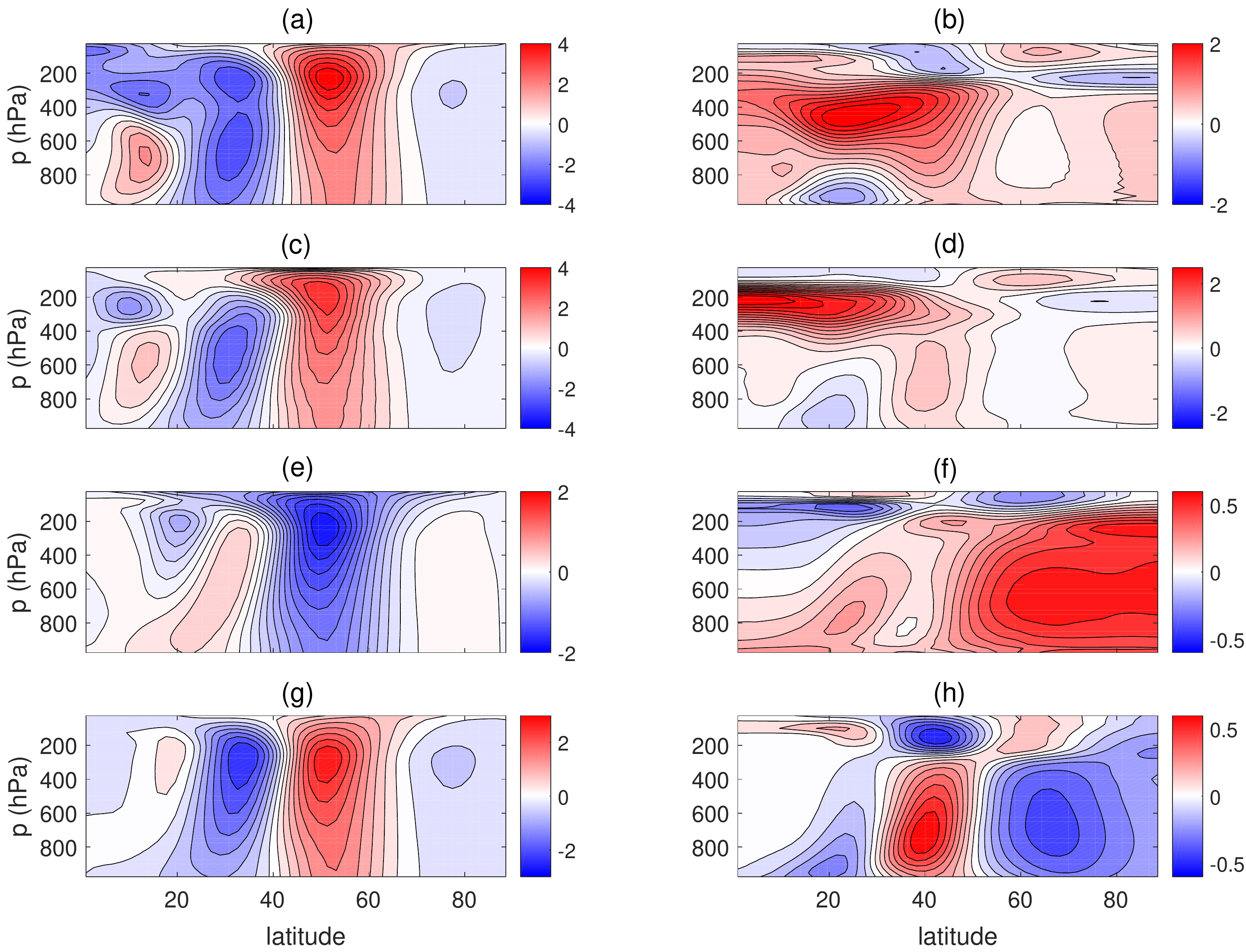}}
\caption{The mean-responses for Tests $1-3$. The left (right) panels show the time-mean zonal-wind in m~s$^{-1}$ (temperature in K). (a)-(b) Test~1: the time-mean response to an imposed Gaussian subtropical thermal forcing $\bar{f} =  0.2 \times \exp{\left[-(p-450)^2/125^2-(|\mu|-25)^2/15^2\right]}$ calculated using an ensemble GCM run (see Fig.~1 in Part~1 for details). The units of $\bar{f}$, $p$, and $\mu$ are K~day$^{-1}$, hPa, and degree, respectively. (c)-(d) similar to (a)-(b) but for tropical forcing $\bar{f} =  0.2 \times \exp{\left[-(p-300)^2/100^2-\mu^2/20^2\right]}$. (e)-(f) Test~2: time-mean response of an ensemble GCM run with Newtonian relaxation timescale that is $10\%$ larger than that of the control-run (see Fig.~2 in Part~1 for details). (g)-(h) Test~3: the first EOF (EOF1) of the control-run, which is the positive phase of Annular Mode (see Fig.~3 in Part~1 for details).}
\label{fig:FDTallf}
\end{figure*}

\begin{figure*}[t]
\centerline{\includegraphics[width=1\textwidth]{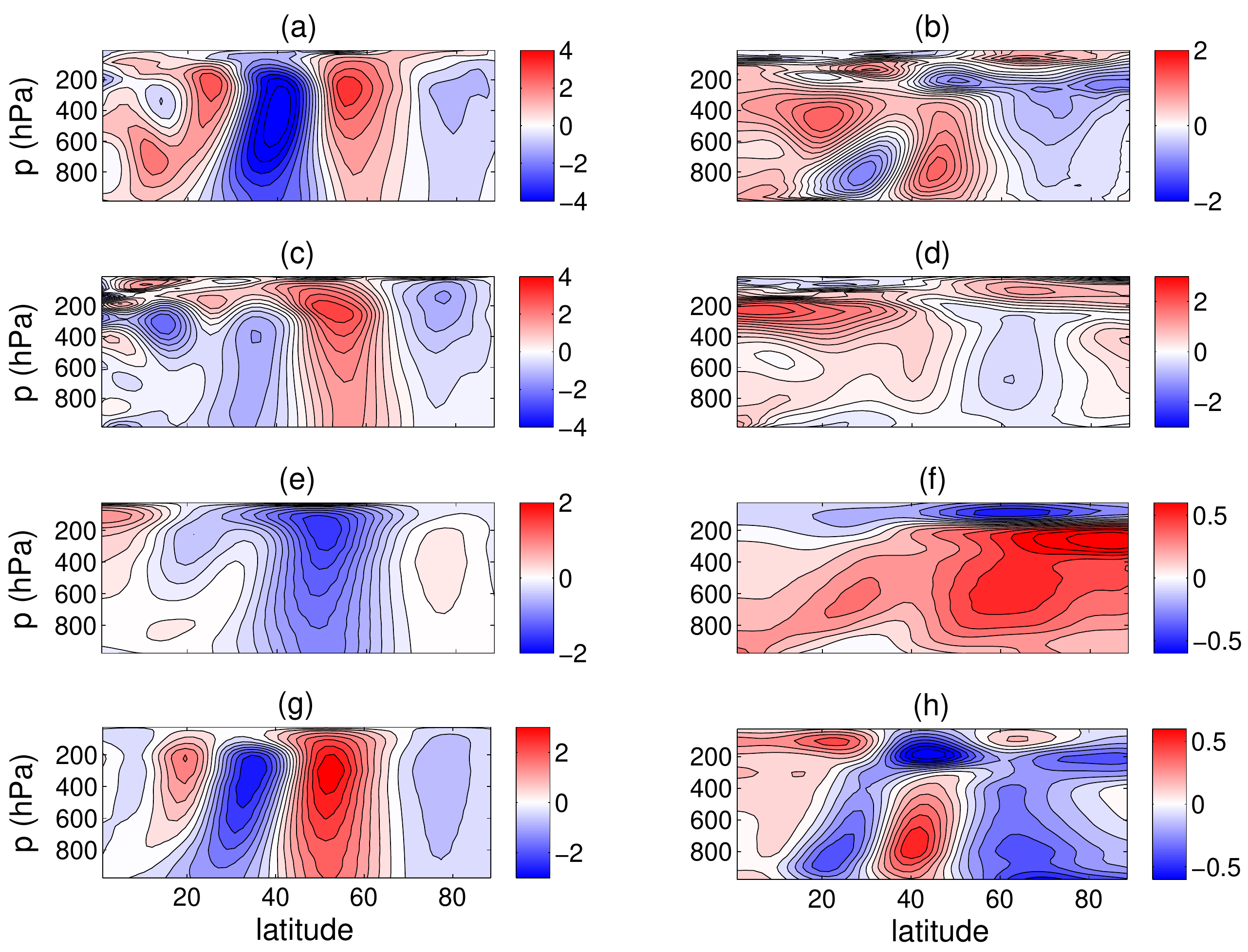}}
\caption{Tests~1-3 for ${\pmb{\mathsf{M}}}_\mathrm{FDT}$. For each Test, the best results for the attempted ranges of $\tau_\infty$ and $n_\mathrm{EOF}$ are shown. The left (right) panels show the time-mean zonal-wind in m~s$^{-1}$ (temperature in K). (a)-(b) Test~1; the relative errors in amplitude are $21 \%$ and $31 \%$, respectively. (c)-(d) similar to Test~1 but for tropical forcing $\bar{f} =  0.2 \times \exp{\left[-(p-300)^2/100^2-\mu^2/20^2\right]}$; the relative errors in amplitude are $6 \%$ and $24 \%$, respectively. Results of (a)-(d) are obtained with $\tau_\infty=30$~days and $n_\mathrm{EOF}=165$. (e)-(f) Test~2;  the relative errors in amplitude are $14 \%$ and $44 \%$, respectively. (g)-(h) Test~3;  the relative errors in amplitude are $13 \%$ and $14 \%$, respectively. Results of (e)-(h) are obtained with $\tau_\infty=30$~days and $n_\mathrm{EOF}=200$.}
\label{fig:FDTall}
\end{figure*}

Tests~1-3 of Part~1 are used to examine the performance of $\tilde{\pmb{\mathsf{M}}}_\mathrm{FDT}$ in predicting the mean-response to external forcings and vice versa. The tests and the true mean-responses are discussed in detail in Part~1 (section~4), but they are also briefly described here and true mean-responses are shown in Fig.~\ref{fig:FDTallf}. For each test of $\tilde{\pmb{\mathsf{M}}}_\mathrm{FDT}$, the best results for the attempted ranges of $\tau_\infty$ and $n_\mathrm{EOF}$ are shown in Fig.~\ref{fig:FDTall} and discussed below.   

In Test~1 we examine the accuracy of $\tilde{\pmb{\mathsf{M}}}_\mathrm{FDT}$ in calculating the time-mean response $\langle \overline{\mathrm{\mathbf{y}}}  \rangle$ to an external subtropical thermal forcing 
\begin{eqnarray}
\bar{f} =  0.2 \times \exp{\left[-(p-450)^2/125^2-(|\mu|-25)^2/15^2\right]}
\label{eq:subt}
\end{eqnarray}
with units of K~day$^{-1}$ (pressure $p$ is in hPa and latitude $\mu$ is in degree). The true response, calculated using an ensemble GCM run forced with this forcing, is shown in Figs.~\ref{fig:FDTallf}(a)-\ref{fig:FDTallf}(b) (see section~4.a of Part~1 for details). The mean-response to this forcing calculated using $\tilde{\pmb{\mathsf{M}}}_\mathrm{FDT}$ is shown in Figs.~\ref{fig:FDTall}(a)-\ref{fig:FDTall}(b). Comparing with the true response (Figs.~\ref{fig:FDTallf}(a)-\ref{fig:FDTallf}(b)) shows that while $\tilde{\pmb{\mathsf{M}}}_\mathrm{FDT}$ can crudely reproduce the patterns of the zonal-wind and temperature response such as the poleward shift of the jet and warming in the subtropical mid-troposphere, it cannot reproduce the amplitude of the response or its patterns at small scales. We have further tested the performance of $\tilde{\pmb{\mathsf{M}}}_\mathrm{FDT}$ using an external tropical forcing 
\begin{eqnarray}
\bar{f} =  0.2 \times \exp{\left[-(p-300)^2/100^2-\mu^2/20^2\right]}.
\label{eq:tro}
\end{eqnarray}
The true mean-response is shown in Figs.~\ref{fig:FDTallf}(c)-\ref{fig:FDTallf}(d). The mean-response calculated using $\tilde{\pmb{\mathsf{M}}}_\mathrm{FDT}$ (Figs.~\ref{fig:FDTall}(c)-\ref{fig:FDTall}(d)) agrees better, both qualitatively and quantitatively, with the true response compared to the subtropical forcing, but there are still notable differences particularly in the zonal-wind.      

The purpose of Tests~2 and 3 is to test whether $\tilde{\pmb{\mathsf{M}}}_\mathrm{FDT}$ can accurately predict the forcing $\langle \overline{\mathrm{\mathbf{f}}}  \rangle$ needed to produce a specified mean-response (i.e., the target). In these tests, for a given target $\langle \overline{\mathrm{\mathbf{y}}}  \rangle$, $\langle \overline{\mathrm{\mathbf{f}}}  \rangle = - \tilde{\pmb{\mathsf{M}}}_\mathrm{FDT} \langle \overline{\mathrm{\mathbf{y}}}  \rangle$ is calculated and applied in the GCM to run a three-member ensemble, where each member is  $45000$~days (note that although we use $\langle \cdot \rangle$ for notation consistency, the forcing $\overline{\mathrm{\mathbf{f}}}$ is time-invariant). The mean-response is then calculated with respect to a three-member control-run ensemble (see section~4 of Part~1 for more details). To minimize the computational cost given the large combination of $(\tau_\infty,n_\mathrm{EOF})$, we have first used $\tilde{\pmb{\mathsf{M}}}_\mathrm{GRF}$, instead of the GCM ensemble run, to test the accuracy of $\overline{\mathrm{\mathbf{f}}}$ calculated using $\tilde{\pmb{\mathsf{M}}}_\mathrm{FDT}$ for the specified targets of Tests~2 and 3. Once the $(\tau_\infty,n_\mathrm{EOF})$ that produces the best results are found for each test, the GCM ensemble run is used to calculate the results that are discussed below and shown in Fig.~\ref{fig:FDTall} (as expected from the results of Part~1, the results of the  GCM ensemble run and $\tilde{\pmb{\mathsf{M}}}_\mathrm{GRF}$ agree well).  

For Test~2, the target is the mean-response to $10\%$ increase in the  Newtonian relaxation timescale of the Held-Suarez setup (Figs.~\ref{fig:FDTallf}(e)-\ref{fig:FDTallf}(f)). As shown in Figs.~\ref{fig:FDTall}(e)-\ref{fig:FDTall}(f), $\tilde{\pmb{\mathsf{M}}}_\mathrm{FDT}$ reproduces the zonal-wind response relatively well both qualitatively and quantitatively except in the tropical stratosphere. The pattern and amplitude of the temperature response, however, are poorly reproduced using $\tilde{\pmb{\mathsf{M}}}_\mathrm{FDT}$. For Test~3, the target is the leading EOF (EOF1) of daily-averaged zonally-averaged anomalous (with respect to the climatology) zonal-wind and temperature, which is the positive phase of the Annular Mode  (Figs.~\ref{fig:FDTallf}(g)-\ref{fig:FDTallf}(h)). As shown in Figs.~\ref{fig:FDTall}(g)-\ref{fig:FDTall}(h), for Test~3 both zonal-wind and temperature responses are reproduced reasonably well using $\tilde{\pmb{\mathsf{M}}}_\mathrm{FDT}$ with the exception of the pattern of the temperature response at high-latitudes. It should be highlighted that in all these tests, the best results are obtained with $\tau_\infty=30$~days, which is around the decorrelation time of EOF1 in the control-run. \citet{gritsun2007climate} and \citet{fuchs2015exploration} also used $\tau_\infty=30$~days, although their GCMs and setups are very different from ours. 
                          
The results of Fig.~\ref{fig:FDTall} show that the LRF calculated using the FDT is relatively accurate for some problems, such as Test~1 with the tropical forcing and Test~3, and inaccurate and only qualitative for some other, such as Test~1 with subtropical forcing and Test~2, consistent with the findings of previous studies \citep[e.g.,][]{gritsun2007climate,lutsko2015applying,fuchs2015exploration}. However, even in Test~3 where the FDT performs the best, some features of the response, such as cooling in the high-latitudes, are poorly reproduced, which can limit applications of $\tilde{\pmb{\mathsf{M}}}_\mathrm{FDT}$, for example for hypothesis-testing. 

The source(s) of the poor performance of $\tilde{\pmb{\mathsf{M}}}_\mathrm{FDT}$ for some problems and its inability to reproduce some patterns of the response is unclear and difficult to identify and might be due to violation of the assumptions underlying Eq.~(\ref{eq:FDT}) such as Gaussianity and/or one or some of the practical issues discussed in section~\ref{sec:fdt}. The departure from Gaussianity in the control-run is found to be substantial in particular for temperature for which the skewness and kurtosis of daily averages can be as large as $2$ and $14$, respectively. Although we use the equivalent of a $990000$-day integration to compute $\tilde{\pmb{\mathsf{M}}}_\mathrm{FDT}$, the limited dataset can certainly still be a source of error. However, the results shown in Fig.~\ref{fig:FDTall} are not substantially better than those obtained using only one fifth of the dataset (but it should be noted that the error in $\tilde{\pmb{\mathsf{M}}}_\mathrm{FDT}$ decreases as $1/\sqrt{N}$ with the length of the dataset $N$ \citep{gritsun2007climate}). Nonlinearity and state-vector reduction are likely not the sources of the poor performance of $\tilde{\pmb{\mathsf{M}}}_\mathrm{FDT}$ given the good performance of $\tilde{\pmb{\mathsf{M}}}_\mathrm{GRF}$ for these tests (see section~4 of Part~1) and the state-vector reduction analysis of Appendix~A in Part~1.                                       
 
The dimension-reduction using projection onto a number of the leading EOFs is another likely source of error. For example, \citet{gritsun2007climate} and \citet{fuchs2015exploration} have found that the LRFs calculated using the FDT perform poorly when the forcings project onto the excluded EOFs. This happens when the forcing pattern differs significantly from the leading EOFs, which is the case for Gaussian forcings. In the following two sections, we show that the dimension-reduction alone can result in a significantly poor performance of ${\pmb{\mathsf{M}}}_\mathrm{FDT}$ for systems with non-normal LRFs, and that the errors increase rapidly with non-normality. This non-normality, not to be confused with non-Gaussianity of the statistics of the state-vector, refers to the non-orthogonality of the eigenvectors of the operator (i.e., the LRF) \citep{farrell1996generalized,farrell1996generalizedII,trefethen1993hydrodynamic,butler1992three,reddy1993pseudospectra} and can result in strong interaction between the components of the forcing and response that project onto the included and excluded EOFs.              

\section{Tests using a linear stochastic equation with $\hat{\pmb{\mathsf{M}}}_\mathrm{GRF}$} \label{sec:fdtSDE}
To focus on the effect of dimension-reduction and eliminate other possible causes for a poor performance of the LRF calculated using FDT, we use a dataset that consists of daily-averaged ${\overline{\mathrm{\mathbf{z}}}}$ obtained from integrating the linear stochastic equation  
\begin{eqnarray}
\dot{\overline{\mathrm{\mathbf{z}}}} = \hat{\pmb{\mathsf{M}}}_\mathrm{GRF} \, \overline{\mathrm{\mathbf{z}}} + \boldsymbol{\zeta}
\label{eq:z}
\end{eqnarray}
using the Euler-–Maruyama method \citep{higham2001algorithmic} with a $0.1$~day timestep for $15$~million days. In (\ref{eq:z}), $\boldsymbol{\zeta}(t)$ is a $200 \times 1$ vector of Gaussian white noise, the $200 \times 200$ matrix $\hat{\pmb{\mathsf{M}}}_\mathrm{GRF}$ is $\tilde{\pmb{\mathsf{M}}}_\mathrm{GRF}$ after inaccurate fast modes are filtered out (see section~3 of Part~1 for details), and ${\overline{\mathrm{\mathbf{z}}}}$ is a $200 \times 1$ state-vector consisting of the coefficients of basis functions
\begin{eqnarray}
\exp \left[-\frac{(|\mu|-\mu_o)^2}{\mu_w^2}-\frac{(p-p_o)^2}{p_w^2} \right]
\label{eq:forcing}
\end{eqnarray}
for $\overline{u}$ and $\overline{T}$ (similar to ${\overline{\mathrm{\mathbf{y}}}}$) where $\mu_w=10^\mathrm{o}$, $\mu_o=0^\mathrm{o}, 10^\mathrm{o}, 20^\mathrm{o}, \dots 90^\mathrm{o}$, $p_w = 75$~hPa, and $p_o=100, 200, 300, \dots 1000$~hPa (see section~3 of Part~1 for details). The advantage of investigating FDT using (\ref{eq:z}) is that while its LRF has the same complexity as that of the GCM, the problem is linear, it produces Gaussian statistics, it can be easily integrated to generate a very long dataset, and the noise is uniformly added to excite all basis functions. As a result of the last two, $\pmb{\mathsf{C}}(\tau)$ can be accurately calculated for large $\tau$, and $\pmb{\mathsf{C}}(0)^{-1}$ and $\left[\int \pmb{\mathsf{C}}(\tau) d\tau \right]^{-1}$ can be computed without the need for dimension-reduction. 

\begin{figure*}[t]
\centerline{\includegraphics[width=1\textwidth]{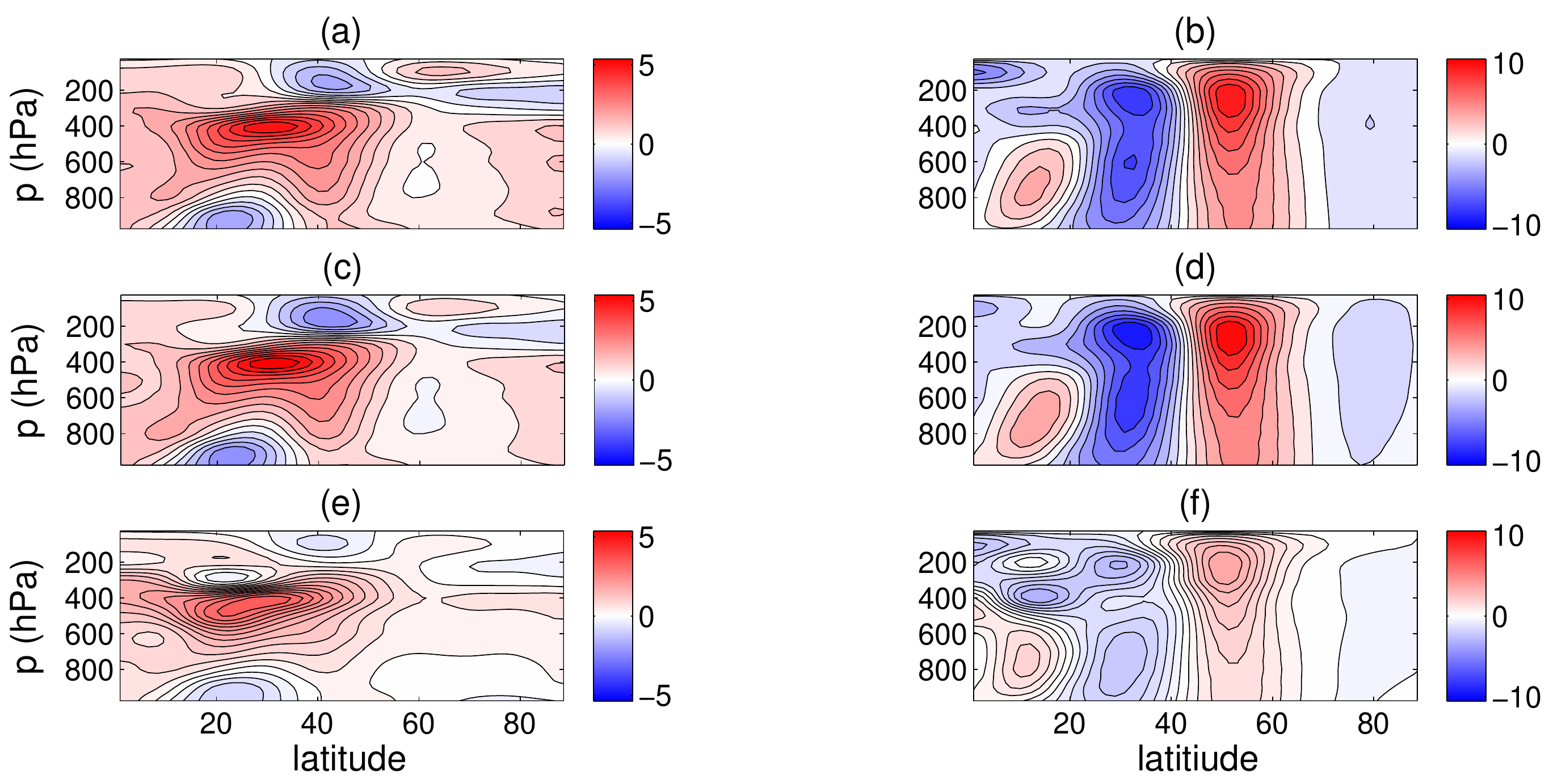}}
\caption{Mean-response to a unit-amplitude thermal forcing of basis function centered at $30^\mathrm{o}$ and $400$~hPa. Left  (right) panels show temperature in K (zonal-wind in m~s$^{-1}$). (a)-(b) the true response calculated using $\hat{\pmb{\mathsf{M}}}_\mathrm{GRF}$. (c)-(d) response calculated using $\tilde{\pmb{\mathsf{M}}}_\mathrm{FDT}$ without dimension reduction. Relative errors in amplitude are $11\%$ (c) and $8\%$ (d). (e)-(f) response calculated using $\tilde{\pmb{\mathsf{M}}}_\mathrm{FDT}$ with dimension reduction using the first $81$ EOFs. Relative errors in amplitude are $31\%$ and $62\%$. If only the component of the response that projects onto the first $81$ EOFs is compared, the relative errors in amplitude are $12\%$ (c), $8\%$ (d), $25\%$ (e), and $64\%$ (f). Results of (c)-(f) are obtained with $\tau_\infty=50$~days.}
\label{fig:SDEfull}
\end{figure*}

Similar to the procedure used previously, ${\overline{\mathrm{\mathbf{z}}}}$ normalized with the standard deviation of each pressure level is used to calculate $\pmb{\mathsf{C}}(\tau)$  for $\tau=0,1, \dots 90$~days, which are then used to compute $\tilde{\pmb{\mathsf{M}}}_\mathrm{FDT}$ from (\ref{eq:FDT}). The mean-responses to a unit-amplitude thermal forcing of the basis function at $(\mu_o,p_o)=(30^\mathrm{o},400~\mathrm{hPa})$ calculated using $\hat{\pmb{\mathsf{M}}}_\mathrm{GRF}$ (the true response) and $\tilde{\pmb{\mathsf{M}}}_\mathrm{FDT}$ (obtained without dimension-reduction) are shown in Figs.~\ref{fig:SDEfull}(a)-\ref{fig:SDEfull}(d). The amplitude and patterns of the two responses agree well. When $\tilde{\pmb{\mathsf{M}}}_\mathrm{FDT}$ is calculated with dimension-reduction using the first $81$~EOFs (which explain $90 \%$ of the variance), the performance of FDT declines significantly (Figs.~\ref{fig:SDEfull}(e)-\ref{fig:SDEfull}(f)) and relative errors in amplitude as large as $60\%$ arise. 

The errors are not simply due to the inability of the dimension-reduced $\tilde{\pmb{\mathsf{M}}}_\mathrm{FDT}$ to capture part of the true response that projects onto the \textit{excluded} EOFs and is forced by the \textit{excluded} component of the forcing (i.e., components of forcing that projects onto the excluded EOFs). In fact similar differences in pattern and errors in amplitude are found if only parts of the responses that project onto the first $81$ EOFs are compared (see the caption). Therefore, the error is to due the inability of the dimension-reduced $\tilde{\pmb{\mathsf{M}}}_\mathrm{FDT}$ to capture part of the true response that projects onto the \textit{included} EOFs and is forced by the \textit{excluded} component of the forcing. This component of the response, which depends on the non-normality of the LRF, can complicate understanding the relationship between the error in the predicted response and the excluded part of the forcing, and can be best understood using simple examples of $2 \times 2$ matrices.                   

\section{Tests using $2 \times 2$ normal and non-normal matrices} \label{sec:fdt2x2}             
\begin{figure*}[t]
\centerline{\includegraphics[width=1\textwidth]{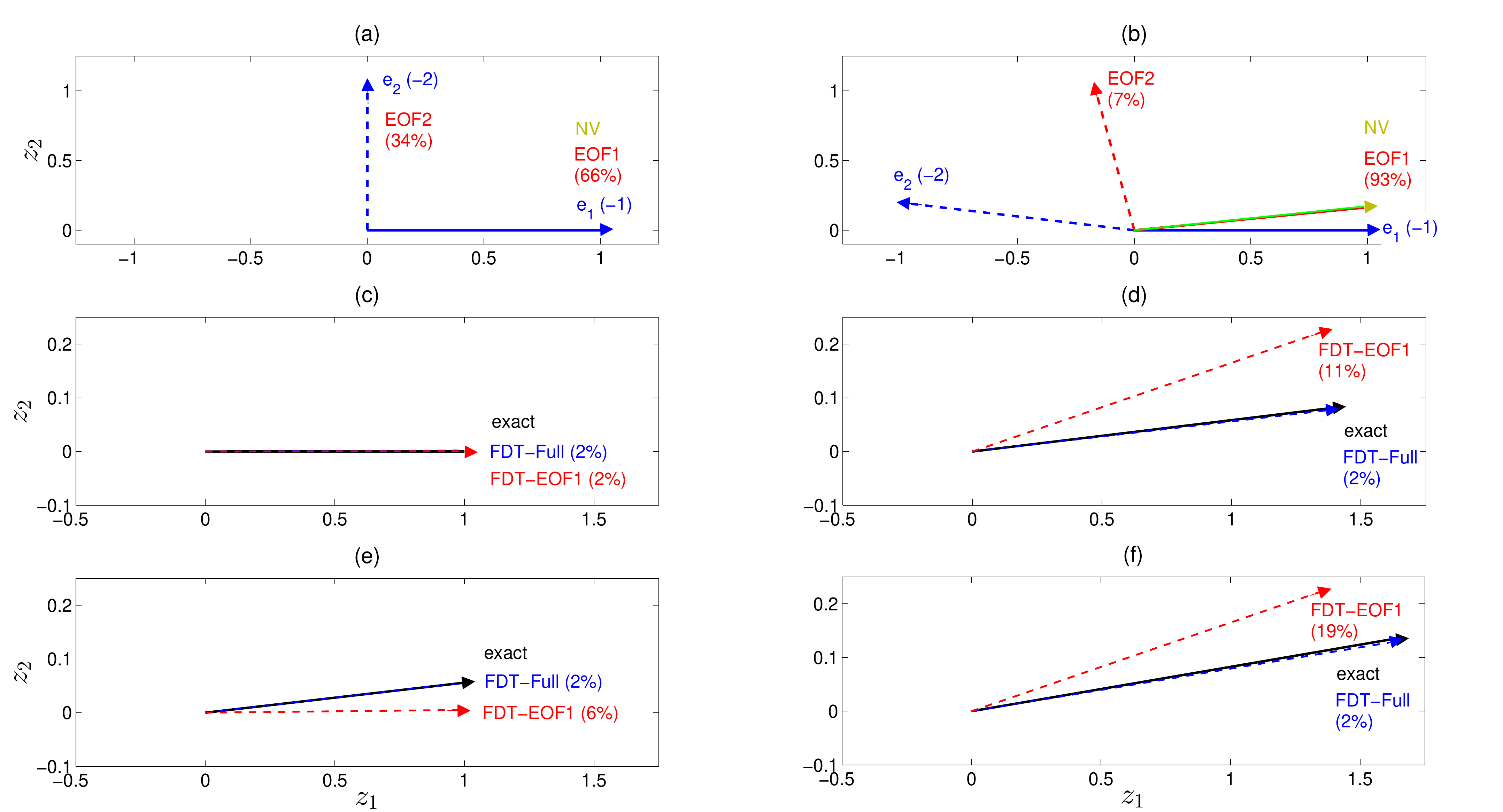}}
\caption{Comparison of the performance of FDT in calculating the mean-response of Eq.~(\ref{eq:2x2}) with normal matrix $\pmb{\mathsf{A}}_\bot$ (left) and non-normal matrix $\pmb{\mathsf{A}}_\angle$ (right). (a)-(b): the blue arrows show the eigenvectors $\mathrm{\mathbf{e}}_1$ and $\mathrm{\mathbf{e}}_2$ with eigenvalues in parentheses; the red arrows show the EOFs with the explained variance in parentheses for $\mathrm{\mathbf{f}}=0$; and the green arrows show the neutral vectors (NV). In (a), $\mathrm{\mathbf{e}}_1$, EOF1, and NV are almost identical (as well as $\mathrm{\mathbf{e_2}}$ and EOF2). In (b), EOF1 and NV are very close. The similarity between EOF1 and NV is expected following the discussion in section~5 of Part~1. (c)-(d): the end of the blue and red arrows show the mean-responses to unit-amplitude forcing $\mathrm{\mathbf{f}}= \mathrm{EOF1}$ obtained, respectively, from full FDT (FDT-Full) and dimension-reduced FDT that only uses EOF1 (FDT-EOF1). The black arrow shows the exact response and the parentheses show the relative error in $\| \cdot \|_2$ (see Eq.~(B3) in Part~1 for definition). (e)-(f) same as (c)-(d) but for unit-amplitude forcing $\mathrm{\mathbf{f}}=(0.9 \times \mathrm{EOF1} + 0.1 \times \mathrm{EOF2})/\| 0.9 \times \mathrm{EOF1} + 0.1 \times \mathrm{EOF2}\|_2$.}
\label{fig:2x2}
\end{figure*}

We focus on a simple linear system
\begin{eqnarray}
\dot{{\mathrm{\mathbf{z}}}} = {\pmb{\mathsf{A}}} \, {\mathrm{\mathbf{z}}} + \boldsymbol{\zeta} + {\mathrm{\mathbf{f}}} 
\label{eq:2x2}
\end{eqnarray}
where ${\mathrm{\mathbf{z}}}=(z_1,z_2)$ is a $2 \times 1$ state-vector and $\boldsymbol{\zeta}(t)$ and ${\mathrm{\mathbf{f}}}$ are $2 \times 1$ vectors of Gaussian white noise and time-invariant external forcing, respectively. To start, we choose ${\pmb{\mathsf{A}}}$ to be either a normal matrix $\pmb{\mathsf{A}}_\bot$ 
\begin{equation}
\pmb{\mathsf{A}}_\bot =  
\begin{bmatrix}
    -1 & 0 \\
    0 & -2
  \end{bmatrix}
\label{eq:Anm}
\end{equation}
or a non-normal matrix $\pmb{\mathsf{A}}_\angle$
\begin{equation}
\pmb{\mathsf{A}}_\angle  =
\begin{bmatrix}
    -1 & 5 \\
    0 & -2
  \end{bmatrix}.
\label{eq:Ann}
\end{equation}
The spectral properties of these matrices are shown in Figs.~\ref{fig:2x2}(a)-\ref{fig:2x2}(b). The matrices have the same eigenvalues $-1$~s$^{-1}$ and $-2$~s$^{-1}$ and the same slowest-decaying eigenvectors $\mathrm{\mathbf{e}}_1$, which are parallel to the horizontal axis. However, while the other eigenvector of $\pmb{\mathsf{A}}_\bot$ is along the vertical axis and hence orthogonal to $\mathrm{\mathbf{e}}_1$, the second eigenvector of $\pmb{\mathsf{A}}_\angle$ is nearly anti-parallel to $\mathrm{\mathbf{e}}_1$ with a $11.4^\mathrm{o}$ angle. As a result, $\pmb{\mathsf{A}}_\angle$ is non-normal, i.e., $\pmb{\mathsf{A}}_\angle \pmb{\mathsf{A}}_\angle^\dag \neq \pmb{\mathsf{A}}_\angle^\dag \pmb{\mathsf{A}}_\angle$, and consequently, in spite of having negative eigenvalues, can lead to non-normal growth in ${{\mathrm{\mathbf{z}}}}$ and instability (see Fig.~1 in \citet{trefethen1991pseudospectra}). Non-normal operators are common in engineering and geophysical/astrophysical flows (see page~579 of \citet{palmer99} for an illustrative example of why) and their significance for the dynamics of the atmosphere and ocean has been recognized through the pioneering papers of Farrell and his colleagues \citep[e.g.,][]{farrell1988optimal,farrell1989optimal,farrell1992adjoint,butler1992three,farrell1996generalized,farrell1996generalizedII} and those of others \citep[e.g.,][]{buizza1995singular,penland1995optimal,ioannou1995nonnormality,zanna2005nonnormal,tziperman2008nonnormal,palmer2013singular}.                   

For both matrices, Eq.~(\ref{eq:2x2}) is integrated for ${\mathrm{\mathbf{f}}}=0$ using the Euler-–Maruyama method with a $0.05$~s timestep for $5$~million seconds. The EOFs of the results are shown in Fig.~\ref{fig:2x2}(a)-\ref{fig:2x2}(b). For the normal matrix, EOF1 and EOF2 are almost identical to $\mathrm{\mathbf{e}}_1$ and $\mathrm{\mathbf{e_2}}$, respectively, while they are different for the non-normal matrix. For both matrices, the results are used to construct $\tilde{\pmb{\mathsf{M}}}_\mathrm{FDT}$ from Eq.~(\ref{eq:FDT}) without dimension-reduction (denoted as FDT-Full) and with only EOF1 retained (denoted as FDT-EOF1). The mean-responses of (\ref{eq:2x2}) with $\pmb{\mathsf{A}}_\bot$ to external forcing ${\mathrm{\mathbf{f}}}=\mathrm{EOF1}$ predicted using FDT-Full and FDT-EOF1 are shown in Figs.~\ref{fig:2x2}(c). Both full and dimension-reduced LRF are very accurate. When the forcing has $10\%$ projection onto EOF2, the LRF calculated using FDT-EOF1 shows a small error, because it cannot capture the part of the response that project onto EOF2 and is forced by the EOF2 component of ${\mathrm{\mathbf{f}}}$. This becomes clear when (\ref{eq:2x2}) is transformed into the EOF-space:  
\begin{eqnarray}
\dot{{\mathrm{\mathbf{a}}}} = \left[\mathrm{\mathbf{EOF}}^{-1} \left({\mathrm{\mathbf{EIG}}} \; \boldsymbol{\Lambda} \; {\mathrm{\mathbf{EIG}}}^{-1} \right)  \mathrm{\mathbf{EOF}} \right]\, {\mathrm{\mathbf{a}}} + \mathrm{\mathbf{EOF}}^{-1} \, {\mathrm{\mathbf{f}}} 
\label{eq:2x2eofs}
\end{eqnarray}                
where ${\mathrm{\mathbf{a}}}=(a_1,a_2)$ is a vector of the coefficients of the EOF1 and EOF2, $\mathrm{\mathbf{EOF}}$ and ${\mathrm{\mathbf{EIG}}}$ are $2 \times 2$ matrices whose columns are the EOFs and eigenvectors of $\pmb{\mathsf{A}}$, respectively, and $\boldsymbol{\Lambda}$ is a diagonal matrix of the eigenvalues of $\pmb{\mathsf{A}}$. The last term is simply a vector of the projections of ${\mathrm{\mathbf{f}}}$ onto the EOFs. The noise term is ignored for convenience. For a normal matrix such as $\pmb{\mathsf{A}}_\bot$, $\mathrm{\mathbf{EOF}}$ and  $\mathrm{\mathbf{EIG}}$ are identical and the first term on the right-hand side reduces to $\boldsymbol{\Lambda} \, {\mathrm{\mathbf{a}}}$. Hence the equations for $a_1$ and $a_2$ decouple. Then if ${\mathrm{\mathbf{f}}}$ does not have any projection on EOF2, $\langle a_2 \rangle =0$ and FDT-EOF1 works as accurately as FDT-Full (Fig.~\ref{fig:2x2}(c)). If ${\mathrm{\mathbf{f}}}$ has a projection onto EOF2, then $\langle a_2 \rangle \ne 0$ causes some errors in FDT-EOF1, which can still accurately calculate $\langle a_1 \rangle$ (Fig.~\ref{fig:2x2}(e)). 

\begin{figure*}[t]
\centerline{\includegraphics[width=0.9\textwidth]{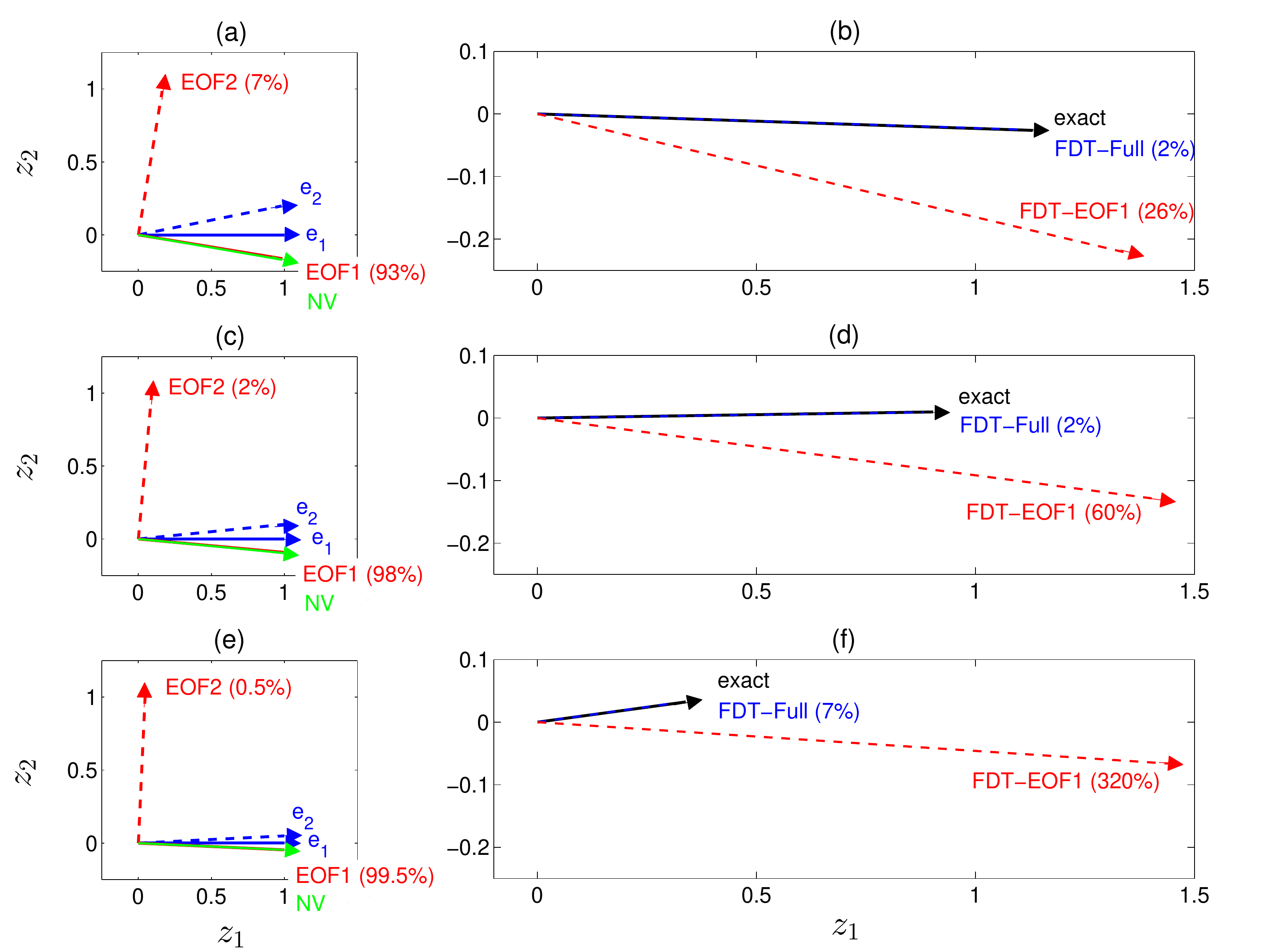}}
\caption{Performance of the dimension-reduced FDT in calculating the mean-response of Eq.~(\ref{eq:2x2}) to unit-amplitude forcing $\mathrm{\mathbf{f}}=(0.9 \times \mathrm{EOF1} + 0.1 \times \mathrm{EOF2})/\| 0.9 \times \mathrm{EOF1} + 0.1 \times \mathrm{EOF2}\|_2$ for three non-normal matrices with different angles between eigenvectors, which are, from top to bottom, $\sim 11.4^\mathrm{o}$, $5.8^\mathrm{o}$, and $2.8^\mathrm{o}$. Left panels: the blue arrows show the eigenvectors $\mathrm{\mathbf{e}}_1$ and $\mathrm{\mathbf{e}}_2$, which have eigenvalues $-1$ and $-2$, respectively; the red arrows show the EOFs with the explained variance in parentheses for $\mathrm{\mathbf{f}}=0$; and the green arrows show the neutral vectors (NV). Right panels: for matrices on the left panels, the end of the blue and red arrows show the responses, respectively, from the full (FDT-Full) and dimension-reduced FDT that only uses EOF1 (FDT-EOF1). The black arrow shows the exact response and the parentheses show the relative error in $\| \cdot \|_2$.}
\label{fig:2x2B}
\end{figure*}

However, for non-normal matrices $\mathrm{\mathbf{EOF}} \ne {\mathrm{\mathbf{EIG}}}$ (Fig.~\ref{fig:2x2}(b)) and the off-diagonal elements of the matrix in the first term on the right-hand side of (\ref{eq:2x2eofs}) can be large, which strongly couples the two equations. In this case, a forcing that only projects onto EOF1 can result in large $\langle a_2 \rangle$, and part of a forcing that projects onto EOF2, even if small, can have a large contribution to $\langle a_1 \rangle$. Neither effects can be captured by FDT-EOF1 which can lead to large errors, as shown in Figs.~\ref{fig:2x2}(d) and \ref{fig:2x2}(f) for $\pmb{\mathsf{A}}_\angle$. The amplitude of these errors depends on the non-normality of $\pmb{\mathsf{A}}$. For example, as shown in Fig.~\ref{fig:2x2B}, for the same forcing ${\mathrm{\mathbf{f}}}$ which has a $10\%$ contribution from EOF2, the error in the response predicted using FDT-EOF1 rapidly increases as the acute angle between the eigenvectors of $\pmb{\mathsf{A}}$ deceases. The decrease in the angle results in a larger off-diagonal term (coefficient of $a_2$) in the equation of $a_1$, which explains the increase in the error. 

\begin{figure*}[t]
\centerline{\includegraphics[width=0.75\textwidth]{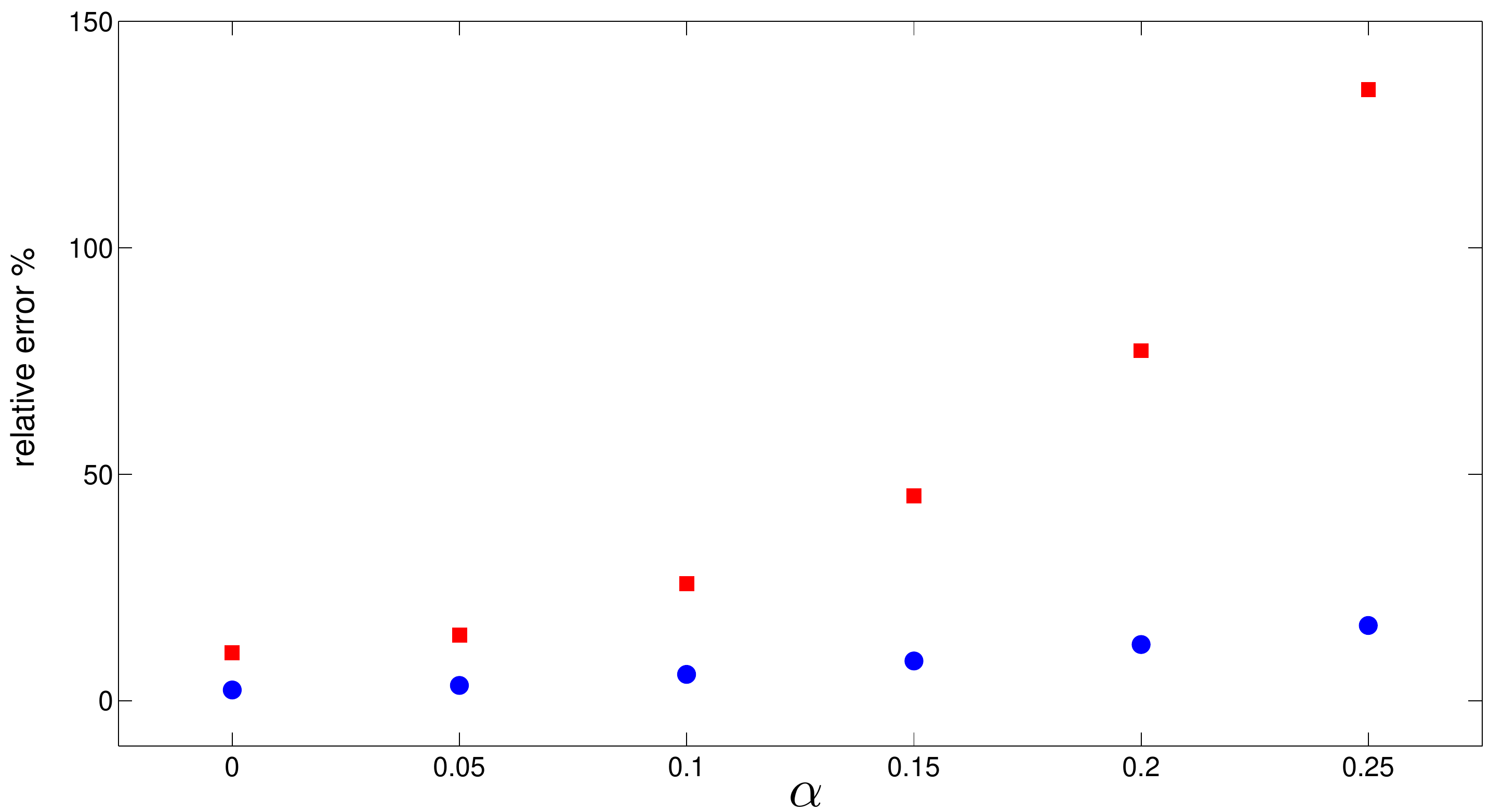}}
\caption{Relative error $\| \langle{\mathrm{\mathbf{z}}}\rangle_\mathrm{FDT-EOF1}-\langle{\mathrm{\mathbf{z}}}\rangle_\mathrm{true} \|_2 \times 100 / \| \langle{\mathrm{\mathbf{z}}}\rangle_\mathrm{true} \|_2$ as a function of $\alpha$ which determines the contribution of EOF2 to forcing ${\mathrm{\mathbf{f}}} = \left[ (1-\alpha) \, \mathrm{EOF1} + \alpha \, \mathrm{EOF2}\right]/\| (1-\alpha) \, \mathrm{EOF1} + \alpha \, \mathrm{EOF2}\|_2$. $\langle{\mathrm{\mathbf{z}}}\rangle_\mathrm{FDT-EOF1}$ is calculated using the LRF constructed from FDT with only EOF1. Blue circles (red squares) show the results for the normal matrix Eq.~(\ref{eq:nor}) (non-normal matrix Eq.~(\ref{eq:nnor})).}
\label{fig:prop}
\end{figure*}

The effect of non-normality complicates understanding the relationship between the error in dimension-reduced FDT predictions and the excluded part of the forcings. For example, non-normality might explain why \citet{fuchs2015exploration} could not find a simple scaling relation between the error in the amplitude of the response and the loss of amplitude of the forcing due to dimension-reduction. While the error in prediction using FDT-EOF1 for a normal matrix is proportional to the percentage of the forcing that projects onto EOF2, the relationship is not linear for a non-normal matrix. This is seen in Fig.~\ref{fig:prop} which shows the relative error of the response that is predicted by the LRF constructed using FDT with only EOF1 retained (FDT-EOF1), as a function of the EOF2 contribution to the forcing for the $2 \times 2$ linear stochastic system (\ref{eq:2x2}) with either normal 
\begin{equation}
\begin{bmatrix}
    -1 & 0 \\
    0 & -2
  \end{bmatrix}
\label{eq:nor}
\end{equation}
or non-normal
\begin{equation}
\begin{bmatrix}
    -1 & -5 \\
    0 & -2
  \end{bmatrix}
\label{eq:nnor}
\end{equation}
operator ${\pmb{\mathsf{A}}}$.

Finally it should be noted that while here we have focused on the effect of non-normality on the errors in the mean-response arising from dimension-reduction, non-normality can also induce errors in the forcing calculated using dimension-reduced FDT for a given mean-response in a similar way.                     

The analyses presented in sections~\ref{sec:fdtSDE} and \ref{sec:fdt2x2} show that dimension-reduction alone can be a substantial source of error in the results obtained from LRFs constructed using FDT because of the non-normality of the system's operator, and demonstrate that the error depends on the relationship between the included and excluded EOFs, eigenvectors of the system's true operator, and the projections of the forcing/response onto the eigenvectors and EOFs. Based on these results, it is likely that errors arising from dimension-reduction are a major, if not the main, contributor to the poor performance of the LRF calculated using FDT in section~\ref{sec:fdttests}. This is further supported by the fact that the best performance of FDT is achieved for Test~3 where the response projects mostly only onto EOF1 (there is little projection onto other EOFs due to the different weightings used in EOF calculations in section~4 of Part~1 and section~\ref{sec:fdttests} of Part~2). 

It should be noted that the problem with non-normality and dimension-reduction discussed here cannot be resolved just by including more EOFs in the LRF construction, because the poorly sampled EOFs degrade the accuracy of LRF, and excluding them is the rationale behind the dimension-reduction strategy. In fact as reported by \citet{fuchs2015exploration} and also found here, including too many EOFs reduces the accuracy of $\tilde{\pmb{\mathsf{M}}}_\mathrm{FDT}$. Furthermore, only focusing on forcings/responses that strongly project onto the leading EOFs is an imperfect solution because it can seriously limit the applications of $\tilde{\pmb{\mathsf{M}}}_\mathrm{FDT}$, as many phenomena of interest do not project onto the natural modes of variability \citep[see, e.g.,][]{scaife2009toward}.  
           
\section{Summary} \label{sec:sum}
In Part~2 of this study, we have calculated the LRF, $\tilde{\pmb{\mathsf{M}}}_\mathrm{FDT}$, for an idealized dry CGM using the FDT. Despite efforts to maximize the performance of the FDT, for example by using a one-million day dataset and trying a range of $(\tau_\infty,n_\mathrm{EOF})$, the $\tilde{\pmb{\mathsf{M}}}_\mathrm{FDT}$ is found to perform poorly for some test cases (section~\ref{sec:model}). To eliminate some of the potential causes of this poor performance, the accurate LRF of this model, $\tilde{\pmb{\mathsf{M}}}_\mathrm{GRF}$, which has been calculated in Part~1 of the paper using Green's functions, is used in a linear stochastic equation driven by Gaussian white noise. Calculating $\tilde{\pmb{\mathsf{M}}}_\mathrm{FDT}$ from very long integrations of this equation reveals that  dimension-reduction by projecting the data onto leading EOFs, which is commonly used for FDT, can significantly degrade the performance of the FDT (section~\ref{sec:fdtSDE}). We show in section~\ref{sec:fdt2x2} that the dimension-reduction causes this error because the LRF of the system is non-normal. For example, as a result of this non-normality, the coefficients of the EOFs can be strongly coupled (see Eq.~(\ref{eq:2x2eofs})), and even small projections of the forcing onto the excluded EOFs can have large contributions to the part of the true response that projects onto the included EOFs. Such contributions cannot be captured by the dimension-reduced $\tilde{\pmb{\mathsf{M}}}_\mathrm{FDT}$ which leads to erroneous predictions.      

The results of this paper point to the operator's non-normality as a major source of difficulty for the practical use of the FDT for the general circulation. The role of non-normality might explain the mixed success and difficulty in understanding some of the results obtained using the FDT in other studies. Given that dimension-reduction is inevitable for calculating LRFs using FDT from limited datasets and that non-normality is common in the oceanic and atmospheric flows, we suggest further investigations of dimension-reduction strategies with a focus on non-normality. 

The results of this paper also provide another example for the applications of the accurate LRF, $\tilde{\pmb{\mathsf{M}}}_\mathrm{GRF}$, that has been calculated in Part~1 of this study (see \citet{pedram15} for another example). Furthermore, we suggest that the linear stochastic equation (\ref{eq:z}) can be helpful in evaluating various strategies related to FDT and in disentangling the contributions of different sources of error because while this equation closely retains the complexity of a full GCM through $\hat{\pmb{\mathsf{M}}}_\mathrm{GRF}$, it is computationally inexpensive to solve and flexible in terms of the driving noise. For example, replacing $\boldsymbol{\zeta}$ with correlated additive and multiplicative noise allows non-Gaussian statistics \citep{sardeshmukh2009reconciling}, which provides a simple framework to investigate the performance of FDT in non-Gaussian systems. 


%


\acknowledgments
We thank Ashkan Borna, Gang Chen, Brian Farrell, Nick Lutsko, Ding Ma, and Saba Pasha for fruitful discussions; Elizabeth Barnes, Packard Chan, Nick Lutsko, Marie McGraw, and Marty Singh for useful comments on the manuscript; and Chris Walker for help with the GCM runs at the initial stage of this study. We are grateful to two anonymous reviewers for insightful feedbacks. This work was supported by a Ziff Environmental Fellowship from the Harvard University Center for the Environment to P.H. and NSF grant AGS-1062016 to Z.K. The simulations were run on Harvard Odyssey cluster. 




\bibliographystyle{ametsoc2014}
\setlength{\bibsep}{2pt plus 0.75ex}
\bibliography{LRFp2_v1}



\end{document}